\documentclass[aip,
amsmath,amssymb,
reprint,%
]{revtex4-1}
\usepackage[english]{babel}
\usepackage{lineno}
\usepackage{graphicx}
\usepackage{dcolumn}
\usepackage{bm}
\usepackage{siunitx}
\usepackage[utf8]{inputenc}
\usepackage[T1]{fontenc}
\usepackage{mathptmx}
\usepackage{etoolbox}
\usepackage{xcolor}
\usepackage{hyperref}
\usepackage{listings}
\usepackage{layouts}

\usepackage[normalem]{ulem}

\DeclareUnicodeCharacter{2212}{\ensuremath{-}}

\makeatletter
\def\@email#1#2{%
 \endgroup
 \patchcmd{\titleblock@produce}
  {\frontmatter@RRAPformat}
  {\frontmatter@RRAPformat{\produce@RRAP{*#1\href{mailto:#2}{#2}}}\frontmatter@RRAPformat}
  {}{}
}%
\makeatother
\begin{document}

\preprint{AIP/123-QED}

\title[Broadband birefringence spectroscopy with sub-kHz precision]{Broadband birefringence spectroscopy with sub-kHz precision}
\author{Maximilian Prinz}
\affiliation{Optical Metrology Group, Faculty of Physics, University of Vienna, Boltzmanngasse 5, 1090 Vienna, Austria}
\affiliation{Vienna Doctoral School in Physics, University of Vienna, Boltzmanngasse 5, 1090 Vienna, Austria}

\author{Dominik Charczun}
\affiliation{Institute of Physics, Faculty of Physics, Astronomy and Informatics, Nicolaus Copernicus University, Grudzi\k adzka 5, PL-87-100 Toru\' n, Poland}

\author{Marcin Bober}
\affiliation{Institute of Physics, Faculty of Physics, Astronomy and Informatics, Nicolaus Copernicus University, Grudzi\k adzka 5, PL-87-100 Toru\' n, Poland}

\author{Mateusz Naro\.znik}
\affiliation{Institute of Physics, Faculty of Physics, Astronomy and Informatics, Nicolaus Copernicus University, Grudzi\k adzka 5, PL-87-100 Toru\' n, Poland}

\author{Piotr Morzy\'nski}
\affiliation{Institute of Physics, Faculty of Physics, Astronomy and Informatics, Nicolaus Copernicus University, Grudzi\k adzka 5, PL-87-100 Toru\' n, Poland}

\author{\\Ulrich Galander}
\affiliation{Optical Metrology Group, Faculty of Physics, University of Vienna, Boltzmanngasse 5, 1090 Vienna, Austria}
\affiliation{Vienna Doctoral School in Physics, University of Vienna, Boltzmanngasse 5, 1090 Vienna, Austria}

\author{Oliver H. Heckl}
\affiliation{Optical Metrology Group, Faculty of Physics, University of Vienna, Boltzmanngasse 5, 1090 Vienna, Austria}

\author{Piotr Mas\l owski}
\email{pima@fizyka.umk.pl}
\affiliation{Institute of Physics, Faculty of Physics, Astronomy and Informatics, Nicolaus Copernicus University, Grudzi\k adzka 5, PL-87-100 Toru\' n, Poland}

\date{\today}

\begin{abstract}
Although current amorphous high-reflective mirror coatings have had tremendous success in metrology applications, they are inherently limited by thermal fluctuations in their coating structure. Alternatively, crystalline coating technology has demonstrated superior thermal noise performance. However, recent studies have revealed birefringent noise sources, raising questions about the limits of frequency stability of high-finesse cryogenic silicon cavities with crystalline mirror coatings.
Here, we show the applicability of cavity-mode dispersion spectroscopy to measure birefringent cavity mode splitting. We measured birefringence induced cavity mode splitting by probing the resonance frequencies of a high-finesse, ultra-low expansion glass cavity with all-crystalline mirror coatings, reaching fractional frequency sensitivity of \SI{5e-14}{} utilizing an optical frequency comb for two orthogonal polarizations. Subsequently, we calculated the static birefringent splitting of the refractive index for \SI{23.8}{\celsius} and \SI{31.3}{\celsius} on the order of \SI{305 \pm 3}{ppm} and \SI{294 \pm 3}{ppm} over \SI{30}{nm} respectively. Furthermore, we propose measurements of dispersive birefringent noise based on optical frequency combs. Our results not only extend the use of optical frequency combs to measure static birefringence, but also implicate a possibility to further study spectrally dependent frequency noise.
\end{abstract}

\maketitle

\section{Introduction\label{sec:introduction}}
Noise sources in high-reflective mirror coatings are the limiting factor for experiments utilizing high-finesse cavities, like interferometric~\cite{abbott_gw150914_2016,punturo_einstein_2010,hall_gravitational-wave_2021, akutsu_overview_2021} and resonant~\cite{naroznik_ultra-stable_2023} gravitational wave detection, space-borne missions such as LISA \cite{colpi_lisa_2024}, optical clocks \cite{oelker_demonstration_2019,kedar_frequency_2023}, cavity-enhanced high-resolution spectroscopy \cite{stankiewicz_cavity-enhanced_2025}, and dark matter experiments \cite{kennedy_precision_2020, nagano_axion_2020, wcislo_new_2018}. Thermal noise within the dielectric coatings has been identified as the main noise source, warranting significant efforts to improve the performance of existing high-reflectivity coatings at lower temperatures or to implement new materials \cite{dickmann_ultra-low_2023} with an overall lower mechanical loss angle. During the last decade, substrate-transferred crystalline coatings based on gallium arsenide / aluminum gallium arsenide (GaAs / AlGaAs) multilayers emerged as viable candidates to fulfill this role and replace conventional dielectric coatings due to their improved thermal noise performance \cite{cole_tenfold_2013, cole_substrate-transferred_2023}.

However, recent studies \cite{yu_excess_2023,kedar_frequency_2023} have revealed power- and polarization-dependent noise sources which limit the performance of these coatings considerably above the thermal noise floor. The existence of birefringence in these coatings was shown and studied as early as their development \cite{cole_tenfold_2013, fleisher_precision_2016, winkler_mid-infrared_2021}. With the rapid onset of newer hybrid coating designs \cite{truong_mid-infrared_2023} and their subsequent comparison in material properties \cite{PernerLukasW.2023Smom,galander_group_2025}, a thorough understanding of static and dynamic birefringent cavity mode splitting and its associated noise becomes essential for further increase in stability and usability of all-crystalline- and hybrid-mirror enhancement cavities.

Here we propose the use of broadband cavity mode dispersion spectroscopy (CMDS) based on an optical frequency comb (OFC) as a tool to further study the effect of birefringence in high-finesse cavities based on crystalline mirrors. We present measurements of dispersive cavity mode shift for s- and p-polarization in a high-finesse, near-infrared cavity based on substrate-transferred all-crystalline mirrors at room temperature (\SI{23.8}{\celsius} and \SI{31.3}{\celsius}). 
We subsequently use the absolute cavity mode frequencies to calculate the static birefringent splitting and determine the underlying change in refractive index. Additionally, we propose the use of comb-based CMDS for measuring the birefringent frequency noise of individual cavity modes.
\section{Theory\label{sec:theory}}

\subsection{Cavity mode positions}
When using highly reflective mirrors in a linear cavity, it is typically assumed that the cavity modes are separated by the free spectral range ($FSR_0$)  given by 

\begin{equation} \label{eq:FSR_linear}
    FSR_0 = \frac{\mathrm{c}}{2 L_\mathrm{cav}},
\end{equation}

with $L_\mathrm{cav}$ being the length of the standing wave cavity and $\mathrm{c}$ the speed of light. However, the resonance condition for cavity mode $q$ with frequency $\nu_q$ is only satisfied for phase shift of integer multiple of $2\pi$ as

\begin{align} \label{eq:resonance_condition}
    \Phi\left(\nu_q\right) = 2\pi \frac{\nu_q}{FSR_0} + \phi_\mathrm{g}\left(\nu_q\right) + \phi_\mathrm{m}\left(\nu_q\right) + \phi_\mathrm{n}\left(\nu_q\right) \overset{!}{=} 2\pi q,
\end{align}

where $\phi_\mathrm{g}\left(\nu_q\right)$ is the frequency dependent Gouy phase shift of the resonator \cite{hoff_tracing_2017}, $\phi_\mathrm{m}\left(\nu_q\right)$ the phase shift introduced by the dispersion of the cavity mirrors, and $\phi_\mathrm{n}\left(\nu_q\right)$ the phase shift due to an intracavity sample, for one round-trip respectively.  At the $q^{\mathrm{th}}$ cavity mode, this additional phase causes a shift $\Delta \nu_q$ in frequency according to

\begin{align} \label{eq:mode_shift}
    \nu_q = \nu_q^0 + \Delta \nu_q = FSR_0\left(q - \frac{\phi_g\left(\nu_q\right)}{2\pi} - \frac{\phi_\mathrm{m}\left(\nu_q\right)}{2\pi} - \frac{\phi_\mathrm{n}\left(\nu_q\right)}{2\pi}\right),
\end{align}

with $\nu_q^0$ being the cavity mode frequency without the added phase. The shift introduced to adjacent cavity modes will be different due to the frequency dependent Gouy phase shift, the dispersive properties of the cavity mirrors or any sample present inside the cavity. By measuring the position of one or more cavity modes and comparing it to a reference grid, the induced phase can be measured. Previously, this formed the basis for the technique of CMDS \cite{cygan_one-dimensional_2015, rutkowski_sensitive_2017, charczun_broadband_2018}, which was developed with focus on the dispersion of intracavity samples rather than mirror coatings.

\subsection{Mirror birefringence from cavity mode positions}
For two incident Gaussian TEM\textsubscript{00} beams  with orthogonal polarizations, the phase shift introduced by the additional terms in Eq. \ref{eq:resonance_condition} will differ if either the intracavity phase shift $\phi_\mathrm{n}$ or the mirror induced phase shift $\phi_\mathrm{m}$ is birefringent. The dispersive Gouy phase $\phi_\mathrm{g}$ depends on the cavity geometry, but is in general negligible compared to the birefringent phase shift added by the mirrors at frequency $\nu_q$.
In the case of an empty cavity with $\phi_\mathrm{n}\left(\nu\right) = 0$ and light resonant at zero angle of incidence, any phase change between two polarization eigenmodes p and s of the incident light stems from a change in the refractive index of the mirror coatings, with a different resonance frequency for p- and s-polarization
\begin{align}\label{eq:pol_dependent_frequency}
    \nu_q^\mathrm{p,s} \approx \nu_q^\mathrm{0}+ \Delta \nu_q^\mathrm{p,s} = FSR_0\left(q - \frac{\phi_\mathrm{m}^\mathrm{p,s}\left(\nu_q\right)}{2\pi} \right),
\end{align}
where $\nu_q^\mathrm{p,s}$ is the resonance frequency of the resonator, and $\Delta \nu_q^\mathrm{p,s}$ the frequency shift introduced by the different mirror phase shifts $\phi_\mathrm{m}^\mathrm{p,s}\left(\nu_q\right)$. The phase difference between the two polarizations at the individual cavity mirrors causes birefringent splitting $\Delta \nu_q^\mathrm{bir}$ of the cavity mode frequency $\nu_q$

\begin{eqnarray}\label{eq:mode_splitting}
\Delta \nu_q^\mathrm{bir} = \nu_q^\mathrm{p} - \nu_q^\mathrm{s} = \Delta \nu_q^\mathrm{p} - \Delta \nu_q^\mathrm{s}
\nonumber 
\\ 
=  -FSR_0 \left(\frac{\phi_\mathrm{m}^\mathrm{p}\left(\nu_q^\mathrm{p}\right)}{2\pi} - \frac{\phi_\mathrm{m}^\mathrm{s}\left(\nu_q^\mathrm{s}\right)}{2\pi}\right)
\nonumber
\\
= -FSR_0 \frac{\Delta\phi_\mathrm{m}^\mathrm{p,s}\left(\nu_q\right)}{2\pi},
\end{eqnarray}

where $\Delta\phi_\mathrm{m}^\mathrm{p,s}\left(\nu_q\right)$ corresponds to the difference in added phase between p- and s-polarization. Note that $\Delta \phi^\mathrm{p,s}_{m}$ is the birefringent phase difference with respect to the optical frequency. Relating this phase difference to the FSR, requires division by the mode number $q$ corresponding to the optical frequency $\nu_q$. By measuring the phase changes at optical frequencies the detection sensitivity compared to the FSR is enhanced by a factor $q$.
\newline
In the case of a linear cavity, the induced phase due to the mirror coatings can be separated into the influence of the individual mirrors M1 and M2 as $\phi_\mathrm{m}\left(\nu_q\right) = \varphi_\mathrm{M1}\left(\nu_q\right) + \varphi_\mathrm{M2}\left(\nu_q\right)$. With this relation, Eq. \ref{eq:mode_splitting} turns into

\begin{equation}\label{eq:mode_splitting_mirror_phase}
   \Delta \nu_q^\mathrm{bir} = - FSR_0 \left( \frac{\Delta\varphi_\mathrm{M1}^\mathrm{p,s}\left(\nu_q\right)}{2\pi} + \frac{\Delta\varphi_\mathrm{M2}^\mathrm{p,s}\left(\nu_q\right)}{2\pi}\right). 
\end{equation}

Here $\Delta \varphi^\mathrm{p,s}_{M1, M2}$ is the difference in added reflected phase between ordinary and extraordinary axis for the two mirrors respectively, assuming that both cavity mirrors' ordinary and extraordinary axes as well as the input polarization are aligned \cite{michimura_effects_2024}.

Finally, the causative frequency-dependent splitting of the refractive index $\Delta n_\mathrm{bir}$ can be calculated from the birefringent splitting $\Delta \nu_\mathrm{bir}$ as given by \cite{yu_excess_2023}
\begin{equation}\label{eq:refractive_index}
    \Delta n_\mathrm{bir} \left(\nu\right)  = \frac{\Delta \nu_\mathrm{bir}\left(\nu \right)}{2 \nu}\frac{L_\mathrm{cav}}{L_\mathrm{pen}\left(\nu\right)}\frac{1}{|\cos{\left( \theta \right)}|},
\end{equation}
where $L_\mathrm{pen}\left(\nu\right)$ is the frequency dependent phase penetration depth of the electric field \cite{babic_analytic_1992} and $|\cos\left(\theta\right)|$ is the correction factor for angular offset in crystal axis orientation between the two cavity mirrors \cite{brandi_measurement_1997}. Following the assumption that the mirrors' axes as well as the input polarization are aligned, the last term in Eq. \ref{eq:refractive_index} becomes 1, due to $\theta=\ang{0}$.

\subsection*{Probing cavity mode positions via Vernier-filtered OFC}
To measure the absolute frequency position of individual cavity modes, a fixed, equidistant reference grid related to $FSR_0$ is needed. The well-defined frequency structure of an OFC lends itself for such a purpose, where the equidistant grid $\nu_q^0 = q \times FSR_0$ in Eq. \ref{eq:mode_shift} is given by the well-known comb Eq.

\begin{align}\label{eq:comb}
    \nu_n = n \times f_\mathrm{rep} + f_0,
\end{align}

where individual comb modes are defined by an integer number $n$, the repetition rate $f_\mathrm{rep}$ and an offset frequency $f_0$.
Translating an OFC over the individual cavity modes by changing either $f_\mathrm{rep}$ or $f_0$, and continuously sampling the comb transmission, creates a spectrum of individual cavity resonances $\nu_q$. If the OFCs $f_\mathrm{rep}$ and the cavities FSR are matched, every cavity mode is sampled, but generally both are detuned from one another, such that

\begin{equation} \label{eq:Vernier}
    k \times f_\mathrm{rep} = l \times FSR_0,
\end{equation}

where $k,l \in \mathbb{N}$. This corresponds to a Vernier filter of the comb by the cavity \cite{gohle_frequency_2007}. A high Vernier filter increases the effective repetition rate, thus increasing the mode spacing such that only every $l^\mathrm{th}$ cavity mode is coincident with the $k^\mathrm{th}$ comb mode.
By referencing the thus sampled cavity mode spectrum to $FSR_0$ we gain not only access to the induced phase $\phi_\mathrm{m}\left(\nu_q\right)$ as in Eq. \ref{eq:mode_shift}, but also the birefringent splitting $\Delta \nu_q^\mathrm{bir}$ in Eq. \ref{eq:mode_splitting} by comparing orthogonal polarizations.

\section{Experiment and Methods}\label{sec:setup_methods}
Full mapping of the cavity mode spectrum and retrieval of the mode positions for two orthogonal polarizations, requires the ability to freely translate the OFC across the transmission spectrum of the cavity, while acquiring transmission spectra at discrete frequency steps. Consequently, we needed an indirect lock, which transferred any drifts of the cavity to the OFC, to guarantee a fixed frequency reference as well as to stay on resonance. For this reason, we directly locked a continuous wave (CW) laser (NKT Photonics ADJUSTIK E15) operating at \SI{1542}{nm} to the cavity. We recorded a heterodyne beat note between the CW laser and the OFC in front of the cavity, locked the resulting beat note $f_\mathrm{beat}$ and changed the setpoint via an acousto-optic modulator (AOM) to move the comb across the cavity resonances. As the CW laser followed the drift of the cavity, the OFC followed the beat note according to the CW laser, thus staying on resonance with the cavity.

\begin{figure*}[htbp]
\centering
\includegraphics[width=\textwidth]{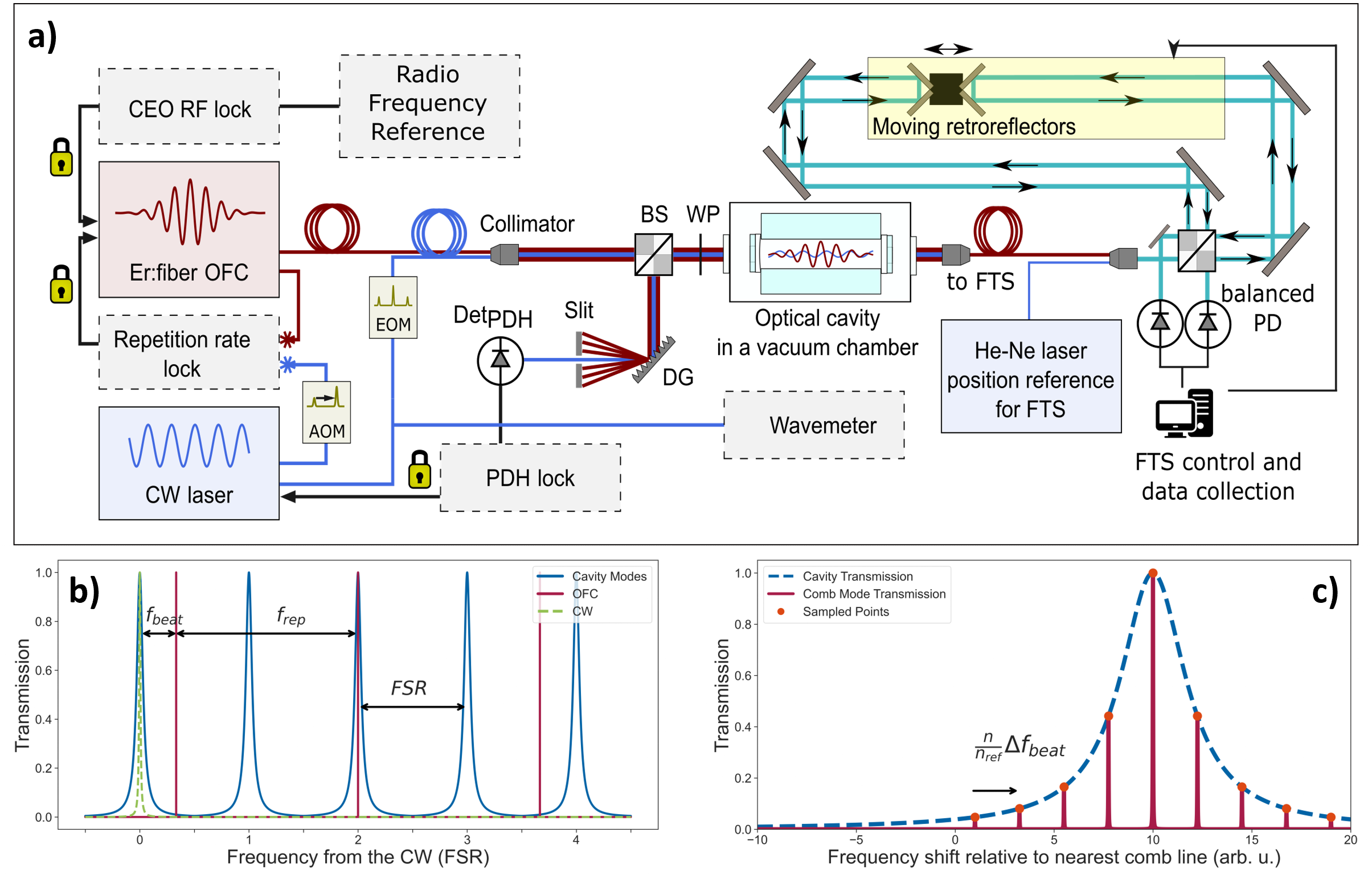}
\caption{a) Schematic of the setup used for CMDS. AOM - acousto optic modulator, EOM - electro optic modulator, BS - beam splitter, DG - diffraction grating, Det\textsubscript{PDH} - Detector for PDH-lock, WP - half-waveplate, FTS - Fourier transform spectrometer, PD - photo detector. b) Locking scheme for CW and OFC. The CW laser (dashed green) is directly locked to a cavity mode, while the OFC (solid red) is directly phase-locked to the CW laser. c) Sampling schematic for an individual cavity mode (dashed blue). The comb mode corresponding to integer $n$ is translated by $\frac{n}{n_\mathrm{ref}}\Delta f_\mathrm{beat}$ when changing the beat note at mode number $n_\mathrm{ref}$. Measuring and interleaving the OFC spectrum (points orange) for each step returns the sampled cavity mode.} 
\label{fig:Fig1_Setup}
\end{figure*} 

\subsection*{General setup and measurement conditions}
The measurement setup is schematically depicted in Fig. \ref{fig:Fig1_Setup} a) and consists of the aforementioned CW laser and an Er:fiber OFC (Menlo Systems) with a nominal repetition rate of \SI{250}{MHz}. Both were coupled to a plano-concave high-finesse cavity which consists of two GaAs/Al\textsubscript{$0.92$}Ga\textsubscript{$0.08$}As crystalline mirrors (Thorlabs Crystalline Mirror Solutions) mounted on an ultra-low expansion glass spacer \cite{Naroznik2022}. The cavity has a length $L_\mathrm{cav}$ of approximately \SI{300}{\milli \metre} and a total loss per mirror of \SI{9.3}{ppm} ($T=$\,\SI{3}{ppm}, $A+S=$\,\SI{6}{ppm}), resulting in a measured finesse of $\mathcal{F} = 340,000$ at \SI{1550}{\nano \metre}. Each mirror consists of 45 layers of AlGaAs and 46 layers of GaAs with design center wavelength of \SI{1556}{nm}, the radius of curvature of the concave mirror is \SI{1}{m}. To determine the actual center wavelength, a transmission matrix model (TMM) \cite{luce_tmm-fast_2022}, based on transmission spectra and layer thickness measurements of a third mirror from the same growth run was calculated. This yielded an actual center wavelength of $\approx$\,\SI{1564}{nm}. The FSR at \SI{1542}{nm} was determined to be \SI{499.6}{MHz}. A set of mode matching lenses ($f_1=f_2=$\,\SI{150}{\milli \metre}) was used to couple the CW and the OFC to the TEM\textsubscript{00} mode of the cavity. The cavity was evacuated to \SI{1e-6}{\milli \bar}, thermally shielded, and temperature stabilized at \SI{23.8 \pm 0.5}{\celsius} and \SI{31.3 \pm 0.5}{\celsius} during the measurements. The output of the cavity was fiber coupled to a custom-built Fourier transform spectrometer (FTS) with sub-nominal resolution \cite{maslowski_surpassing_2016, rutkowski_optical_2018} where the cavity transmission per comb mode was measured with a custom-built auto-balanced detector.

\subsection*{Direct lock of CW laser to cavity}
The frequency of the CW laser was modulated via a fiber EOM (iXblue MPX-LN-0.1) at \SI{10}{MHz} and locked to a cavity mode via the Pound-Drever-Hall (PDH) locking scheme \cite{black_introduction_2001}. 
We used a high-bandwidth servo controller (Toptica FALC 110) to feed back to the CW laser. High-frequency noise was suppressed by modulating the laser frequency with an AOM (AA Optoelectronic MT80-IIR30-Fio-PM0.5-J1-A-Ic2), while internal thermal tuning compensated for slow drifts. To increase the signal-to-noise ratio (SNR) of the error signal and avoid saturation at the detector (Thorlabs APD410C/M) for the PDH-lock, a grating was used to separate the CW laser from the OFC. In order to control the polarization of the CW laser and the OFC in front of the cavity, a polarimeter (Thorlabs PAN5710IR3) was used to set the polarization either to s or p with respect to the laboratory reference plane. Due to that, the lock between the cavity and the CW laser had to be broken between the two measurements. To ensure comparability between the measurements, we locked the CW laser to the same cavity mode after switching polarizations, by comparing the measured beat note, repetition rate and carrier-envelope offset frequency of the OFC to the previous values.

\subsection*{Indirect lock of OFC to cavity}
The repetition rate $f_\mathrm{rep}$ of the OFC was stabilized by locking the beat note frequency between the OFC mode $n_\mathrm{ref}=774541$ and the CW laser to a value close to $\approx$\,\SI{63}{MHz}. The carrier-envelope offset frequency  $f_0$ was locked directly to a hydrogen maser signal available in the laboratory via a fiber link to the Space Research Centre of Polish Academy of Sciences \cite{sliwczynski_dissemination_2013,morzynski_absolute_2015}. This signal also served as a reference for all frequency measurements.   The total power of the OFC together with the CW laser was measured to be $\approx$\,\SI{20}{\milli \watt} in front of the cavity. 
\newline
Fig. \ref{fig:Fig1_Setup} b) shows the indirect lock of the OFC to the cavity. The CW laser (dashed green) was locked to one of the  cavity modes (solid blue), while the beat note between the CW and the nearest comb mode was used to lock $f_\mathrm{rep}$ of the OFC. The cavity acted as a Vernier filter for the comb, due to the difference in repetition rate and FSR as given in Eq. \ref{eq:Vernier}. In our case, every $197^\mathrm{th}$ comb mode overlapped with every $99^\mathrm{th}$ cavity mode, resulting in an effective FSR of $\approx$\,\SI{49.45}{\giga \hertz}. As the cavity length was not actively stabilized, any length drifts were translated to the CW laser which in turn translated the drift to the OFC's $f_\mathrm{rep}$, enabling relative measurements of cavity modes.

\subsection*{Sampling procedure}
The cavity modes were sampled corresponding to Fig. \ref{fig:Fig1_Setup} c). By changing the beat note $f_\mathrm{beat}$ between the CW laser and the reference comb mode $n_\mathrm{ref}$ by $\Delta f_\mathrm{beat}$, we translated the $n^{\mathrm{th}}$ comb mode relative to the reference cavity mode by a value $\frac{n}{n_\mathrm{ref}}\Delta f_\mathrm{beat}$. For each value of $f_\mathrm{beat}$ we measured the transmission signal in the sub-nominal FTS. We corrected for the drift of the cavity during each measurement, by tracking $f_\mathrm{rep}$, $f_0$ and $f_\mathrm{beat}$ for each interferogram and calculating the drift of the CW laser compared to the first measurement. Finally, we interleaved the comb mode spectra in order to retrieve the Vernier-filtered cavity mode spectra. As depicted in Fig. \ref{fig:Fig1_Setup} c) the thus measured cavity mode positions were shifted from the theoretical value by the dispersive shift in Eq. \ref{eq:pol_dependent_frequency}.

\begin{figure*}[htbp]
\centering
\includegraphics[width=1\textwidth]{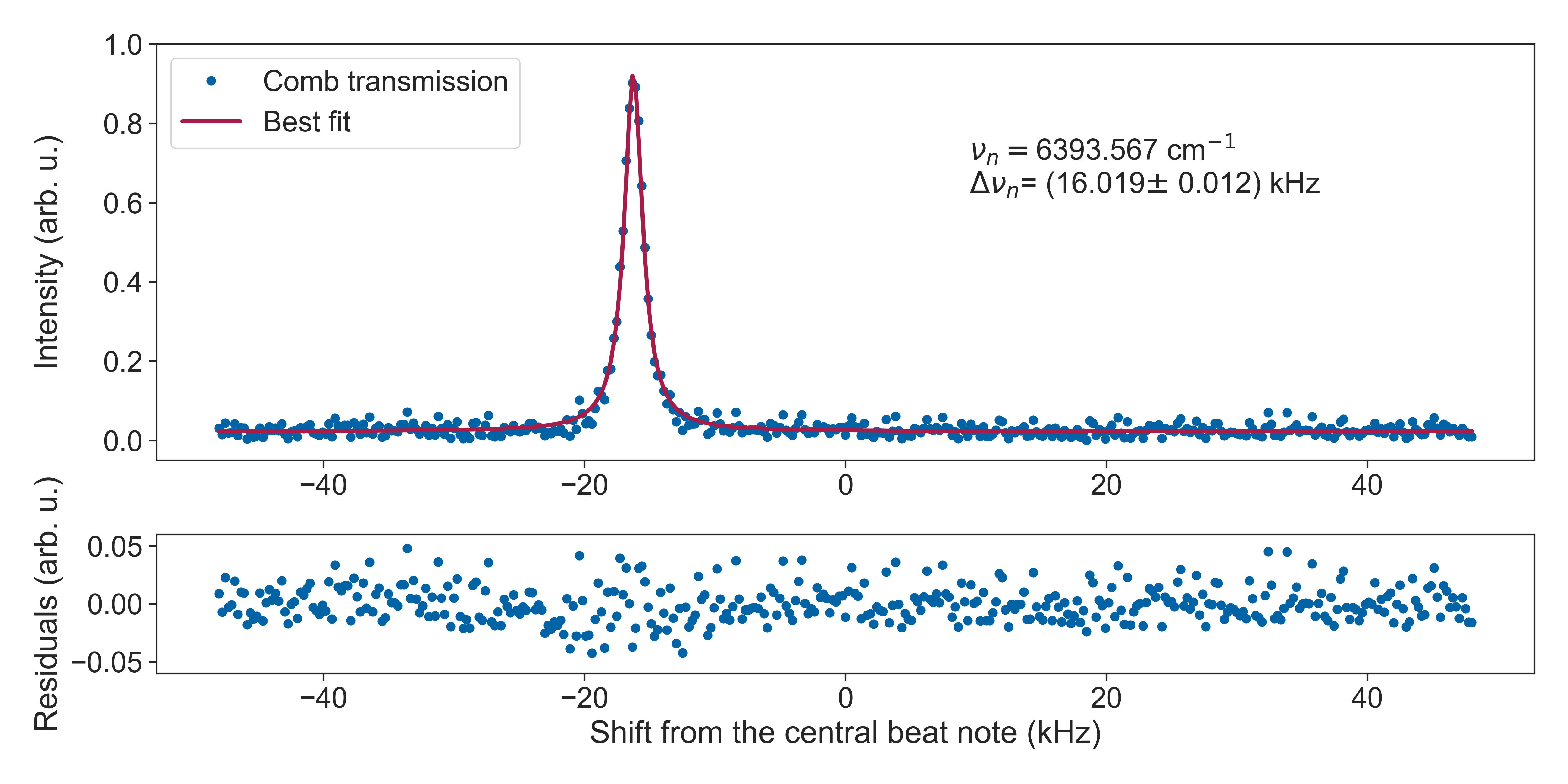}
\caption{Exemplary data for p-polarization at \SI{23.8}{\celsius}. Sampled cavity mode with Lorentzian fit and residuals, relative to the central beat note of the OFC and the CW laser. The central beat note was chosen, such as to maximize the transmission of the OFC through the cavity. The baseline correction of the fit is not included.}
\label{fig:Fig2_Mode_Spectrum}
\end{figure*} 

An example set of measurements for p-polarization at \SI{23.8}{\celsius} is presented in Fig. \ref{fig:Fig2_Mode_Spectrum}, which shows a completely sampled cavity mode spectrum, where each point corresponds to a different value of $f_\mathrm{beat}$ for the comb mode. A Lorentzian function was fitted to the data and its line parameters were retrieved. With the thus extracted center frequency the cavity mode shift was determined to be \SI{16.019 \pm 0.012}{\kHz} from the expected position of $\nu_\mathrm{n} = \SI{6393.567}{\per \cm}$. The residuals depicted below are reasonably flat with root-mean-squared-error (RMSE) of $0.00024$ and depict no obvious deviation from the Lorentzian line-shape. The mean RMSE for all fit-residuals was calculated as \SI{0.0016 \pm 0.0095}{}.

\subsection*{Penetration depth simulation}
To calculate $\Delta n_\mathrm{bir}$ as in Eq. \ref{eq:refractive_index}, knowledge about the frequency dependent phase penetration depth $L_\mathrm{pen}\left(\nu\right)$ is required. We retrieved the independent data from a similar measurement to Ref. \onlinecite{perner_simultaneous_2023}. For this, we measured the broadband transmission of a third mirror from the same growth run in the same custom-built white light interferometer as in Ref. \onlinecite{galander_group_2025}. Subsequently, we cleaved the sample and measured the layer thicknesses in a scanning electron microscope (Zeiss Supra 55 VP). We fitted the measured transmission using a TMM \cite{luce_tmm-fast_2022}, where we fed the retrieved layer thicknesses and refractive index values for GaAs and Al\textsubscript{$0.92$}Ga\textsubscript{$0.08$}As \cite{afromowitz_refractive_1974} together with a global layer scaling parameter to compensate uncertainty in the layer thickness. From the resulting complex reflection spectrum of a single mirror we extracted the phase, differentiated it with respect to angular frequency, and calculated the phase penetration depth, based on Eq. (4) and (26) in Ref. \onlinecite{babic_analytic_1992}.

\section{Results}\label{sec:results}
\begin{figure*}[htbp]
\centering
\includegraphics[width=1\textwidth]{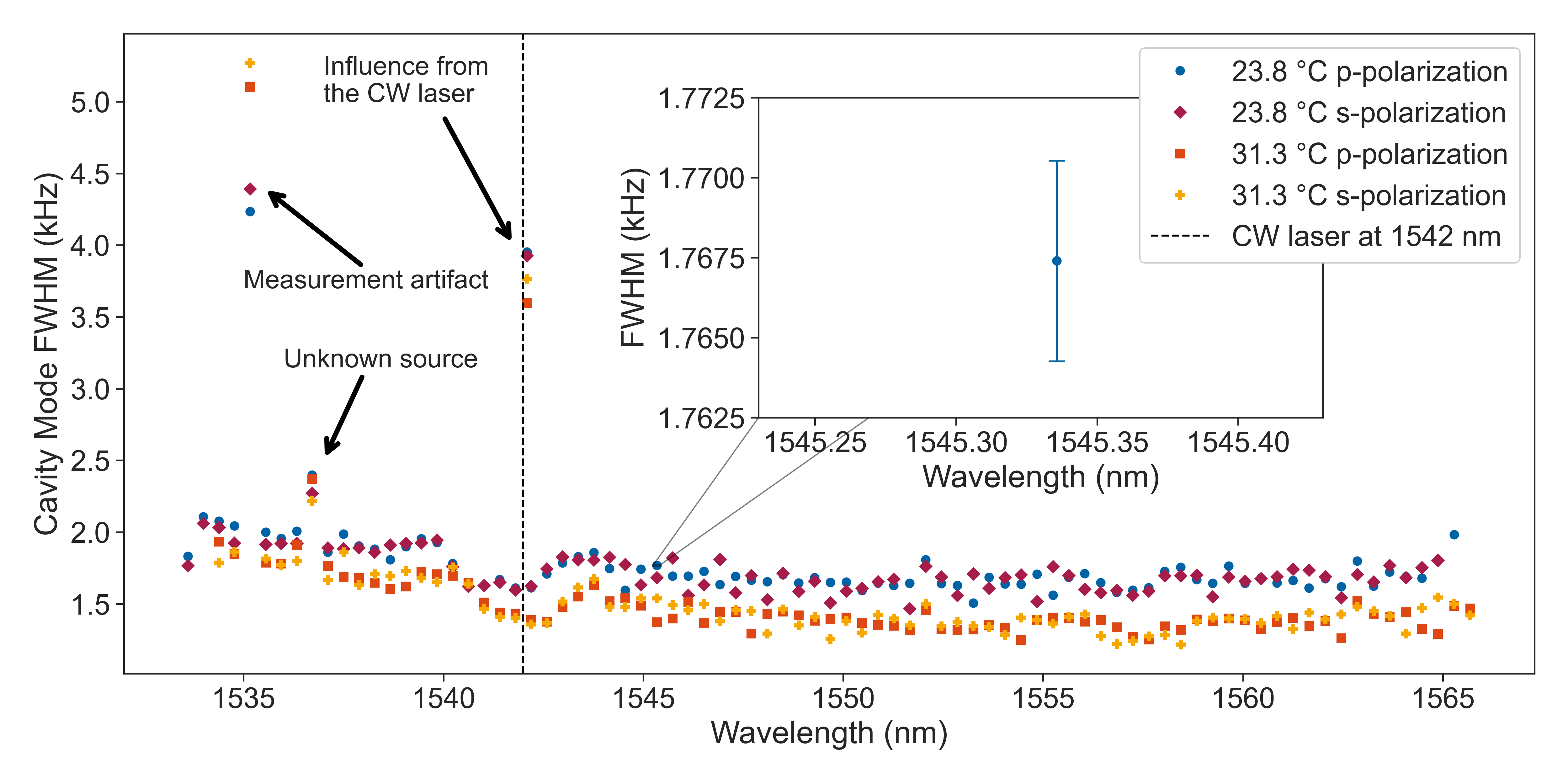}
\caption{Cavity mode width spectrum from Lorentzian fits, the inset shows the one-sigma uncertainty of the fitted mode width. The arrows depict outliers in the mode spectrum, which have increased FWHM. While the leftmost outlier could be linked to a systematic measurement artifact where the mode spectrum was distorted, the outliers at \SI{1536.7}{nm} show genuinely higher mode width, suggesting residual absorption inside the cavity. The right most outliers stem from spectral overlap with the CW laser, leaking into the measurement.}
\label{fig:Fig_FWHM}
\end{figure*}

\begin{figure*}[htbp]
\centering
\includegraphics[width=1\textwidth]{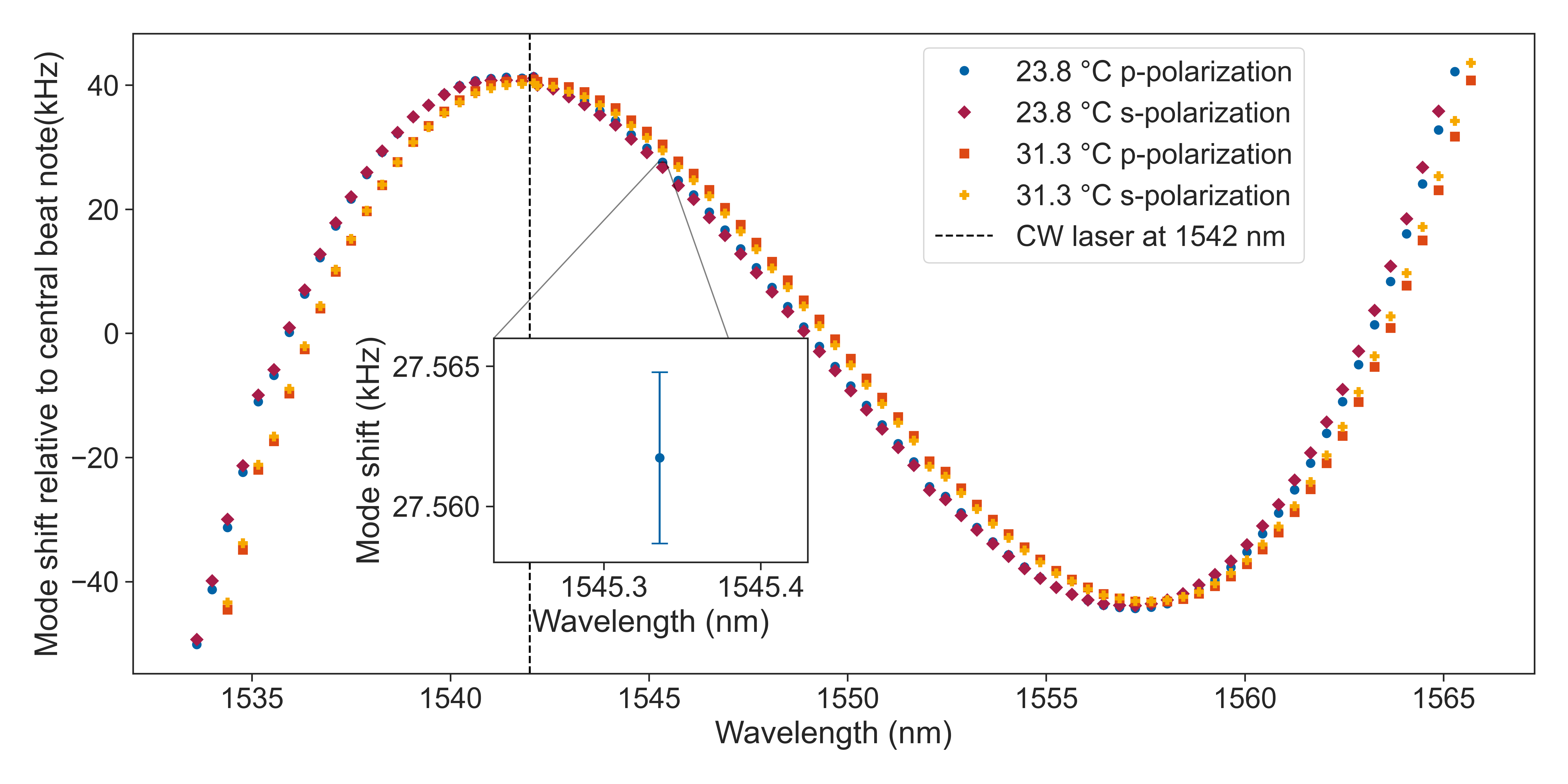}
\caption{Dispersive mode shift for two polarization states compared to the optical frequency comb. The inset shows the one-sigma uncertainty in the measured mode shift. Note that all curves cross each other at the wavelength of the CW laser close to \SI{1542}{nm}. While the FWHM is distorted for some measurements, the retrieved mode position is robust against measurement artifacts and outliers.}
\label{fig:Fig3_Mode_shift}
\end{figure*} 

\begin{figure*}[htbp]
\centering
\includegraphics[width=1\textwidth]{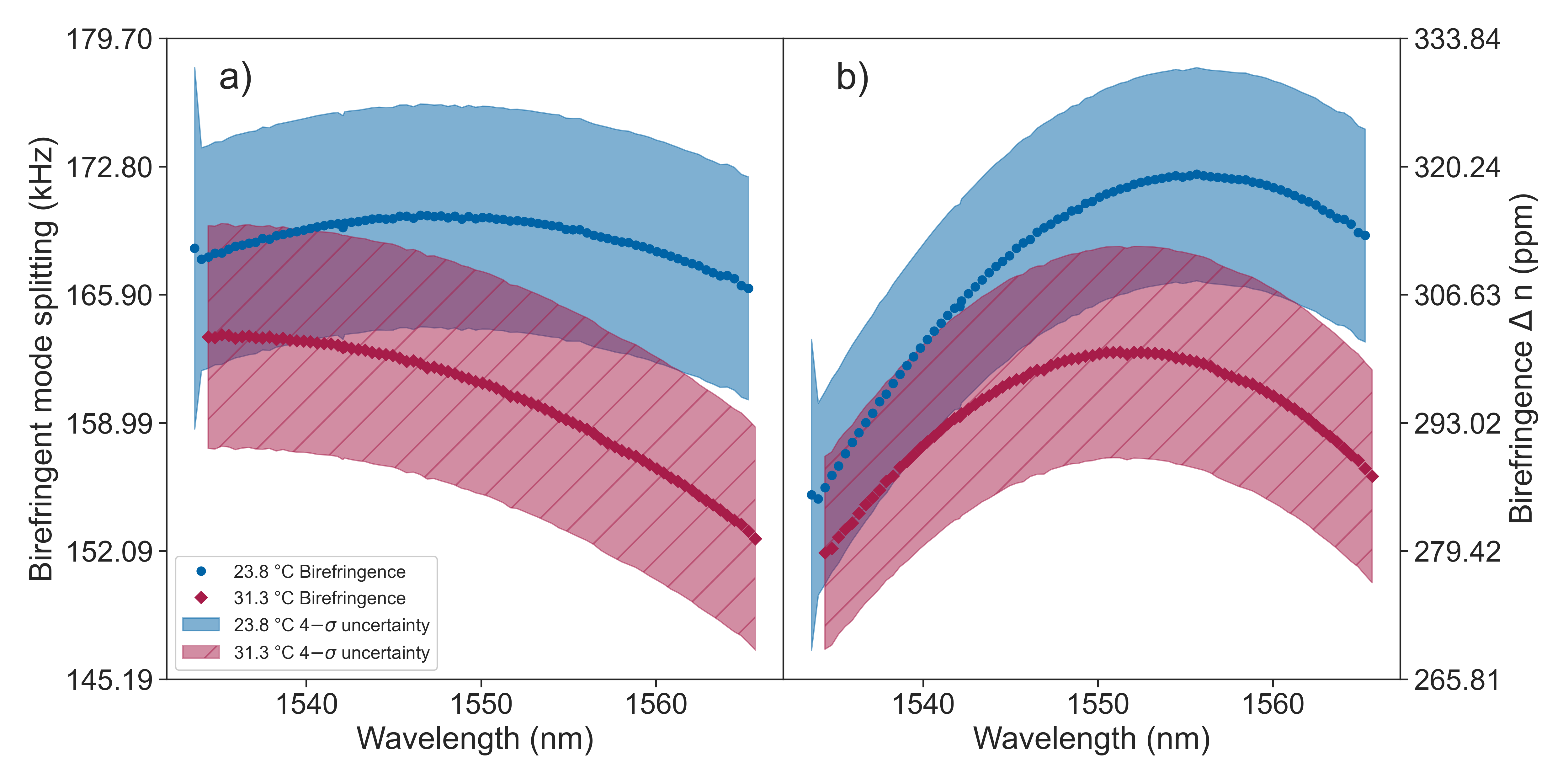}
\caption{ a) Static birefringent cavity mode splitting after correcting for cavity drifts, b) resulting change in refractive index. The shaded areas correspond to the 4-sigma uncertainty of each measurement. Note that the increased uncertainty for the lowest wavelength at \SI{23.8}{\celsius} stems from the fit, as the cavity mode was close to outside the sampled range.}
\label{fig:Fig4_Birefringence}
\end{figure*}

Using the above schemes we recorded a total of 326 cavity modes (82 per polarization at \SI{23.8}{\celsius}, and 81 per polarization at \SI{31.3}{\celsius}), each sampled with 401 beat notes, over a spectral span of approximately \SI{32}{nm} for both temperatures. We tuned $f_\mathrm{beat}$ by \SI{96.240}{kHz}, which resulted in discrete frequency steps of \SI{240}{Hz} for comb mode $n_\mathrm{ref}=774541$ closest to the CW laser. We inferred the one-sigma frequency uncertainties in terms of mode position and width from the tracked frequency precision of $f_\mathrm{rep}$, $f_0$ and $f_\mathrm{beat}$, as well as the fit uncertainty. We note that the uncertainty associated to the fits outweighs the uncertainty of $f_\mathrm{rep}$, $f_0$ and $f_\mathrm{beat}$ by two orders of magnitude.
Fig. \ref{fig:Fig_FWHM} shows the retrieved cavity mode width spectrum, based on Lorentzian fits, which resulted in mean mode FWHM values of \SI{1.798 \pm 0.009}{kHz} and \SI{1.783 \pm 0.007}{kHz} for p- and s-polarization at \SI{23.8}{\celsius}, as well as \SI{1.543 \pm 0.005}{kHz} and \SI{1.552 \pm 0.005}{kHz} for p- and s-polarization at \SI{31.3}{\celsius} including outliers. While the outliers at $\approx \SI{1542}{nm}$ can be explained by residual influence of the reference CW laser in the comb spectrum due to spectral overlap, the origin of outliers at $\approx \SI{1536.7}{nm}$ remain unknown, while those at $\approx \SI{1535.2}{nm}$ can be linked to a measurement artifact that distorted the mode spectrum. 
The resulting dispersive mode shift relative to the nearest comb mode for s- and p-polarization is shown in Fig. \ref{fig:Fig3_Mode_shift}. Combining this shift with the absolute frequency axis defined by the OFC, and normalizing any cavity drift to the central beat note, we were able to calculate the absolute optical frequency of each individual cavity mode. In order to correct for any drifts that occurred while changing the polarization, we corrected the difference between absolute cavity mode frequencies, such that the birefringent splitting at the frequency of the CW laser equals a previously determined value of \SI{169.8 \pm 1.5}{kHz} for \SI{23.8}{\celsius} and \SI{163 \pm 2}{kHz} for \SI{31.3}{\celsius}. The resulting frequency difference corresponding to the birefringent mode splitting and the resulting difference in refractive index as calculated with Eq. \ref{eq:refractive_index} is visible in Fig. \ref{fig:Fig4_Birefringence} a) and b). By comparing Fig. \ref{fig:Fig_FWHM} with Fig. \ref{fig:Fig4_Birefringence} a) we see that the birefringent mode splitting is two orders of magnitude larger than the cavity mode FWHM. Ultimately, by evaluating Eq. \ref{eq:refractive_index} at a wavelength of \SI{1542.09}{nm}, we report birefringence values of the crystalline cavity of \SI{305 \pm 3}{ppm} for \SI{23.8}{\celsius} and \SI{294 \pm 3}{ppm} for \SI{31.3}{\celsius}.

\section{Discussion}\label{sec:disc}
\subsection*{Static birefringence}
Since no significant difference was found between the mode widths of the probed polarizations we can not infer a "preferred" polarization for the crystalline mirrors, which sees lower losses (and thus narrower FWHM) compared to the other polarization. This is in agreement with a previous study that showed no difference between transmission and losses between polarization eigenmodes \cite{kedar_frequency_2023}. The resulting static birefringent splitting is lower than previous reports, as Ref. \onlinecite{yu_excess_2023} reported a birefringent mode shift of $\Delta\nu_\mathrm{bir}\approx \SI{200}{kHz}$ and a resulting birefringent difference of $\Delta n_\mathrm{bir} = \SI{690 \pm 3}{ppm}$ and $\SI{792\pm 2}{ppm}$ at \SI{1542}{nm} respectively. We were not able to determine the relative angular separation between the (100) crystal axis of the mirrors, and assumed the case $\theta=\ang{0}$ in Eq. \ref{eq:refractive_index}, as well as $L_\mathrm{cav}=\SI{30}{cm}$. As $L_\mathrm{pen}\left(\nu\right)$ strongly depends on the choice of refractive index model, layer thicknesses and unknown residual absorption, the frequency dependence given in Fig. \ref{fig:Fig4_Birefringence} b) can change significantly, depending on the model choice and measurement parameter information. We employed a semi-empirical model that neglects the extinction coefficients of GaAs/AlGaAs, see Ref. \onlinecite{afromowitz_refractive_1974}.

The uncertainties associated with our results for $\Delta \nu_\mathrm{bir}$ are dominated by the drift compensation between measurements, based on the previously determined birefringent splitting at the CW laser, where reliable data of $f_\mathrm{rep}$ and $f_\mathrm{0}$ are lacking. The presented uncertainties correspond to the accuracy of the measurement, meaning that the measured behavior of $\Delta \nu_\mathrm{bir}$ is highly-precise, but the exact values can be shifted vertically within the uncertainty-bounds. It is self-evident, that future experiments should simultaneously interrogate the cavity polarization eigenmodes, similar to Ref. \onlinecite{yu_excess_2023} to overcome this limitation. The uncertainties associated to $\Delta n$ are retrieved via error propagation of the $1-\sigma$ uncertainties of $\Delta \nu_\mathrm{bir}$. Due to the unknown relative angular separation $\theta$, the curves can be shifted vertically for both temperatures. We strongly emphasize that a more accurate determination of $\Delta n\left(\nu\right)$ depends on rigorous characterization of the cavity mirrors, in terms of wavelength dependent total losses, layer thicknesses as well as sophisticated complex refractive index models.

The different behavior for the two temperatures in Fig. \ref{fig:Fig4_Birefringence} suggests a temperature dependence with higher rate of change for higher temperature. By moving away from the temperature of zero-thermal-expansion (below \SI{10}{\celsius} for the here investigated cavity) the FSR becomes more prone to drifts, while additional temperature-dependent data on the birefringence of the crystalline mirrors is not available. Note that at $\approx \SI{1565}{nm}$, which corresponds to the assumed center wavelength of the mirror pair used in the cavity, the curves diverge highly. Since only one half of the mirror stopband was accessed, no definite statement can be made with regards to the overall behavior of the static birefringence. 

In principle the lowest detectable change in refractive index for one cavity mode is given by the precision of the mode position, which in our measurements was on the order of \SI{10}{Hz} at \SI{194}{THz} which results in noise equivalent fractional sensitivity of $\Delta\nu / \nu \approx\SI{5e-14}{}$, similar to previously reported levels of optical anisotropy measurements \cite{bailly_highly_2010}. We assume that with a tighter control loop, more sophisticated isolation from environmental perturbation and temperature stabilization, we could further improve this sensitivity. Due to the drift between the measurements and the thus increased uncertainty for the birefringent mode splitting, the fractional frequency stability increased to $\Delta\nu / \nu \approx \SI{1e-11}{}$ for the here presented values of $\Delta \nu_\mathrm{bir}$. Simultaneous probing of both polarizations in the future will overcome this limitation, and fully utilize the available sensitivity from CMDS. The idea of such a measurement scheme based on dual-comb spectroscopy to overcome the described limitations is shown below. 

\subsection*{Dispersive birefringent noise - suggested experiment}
\begin{figure*}[htbp]
\centering
\includegraphics[width=1\textwidth]{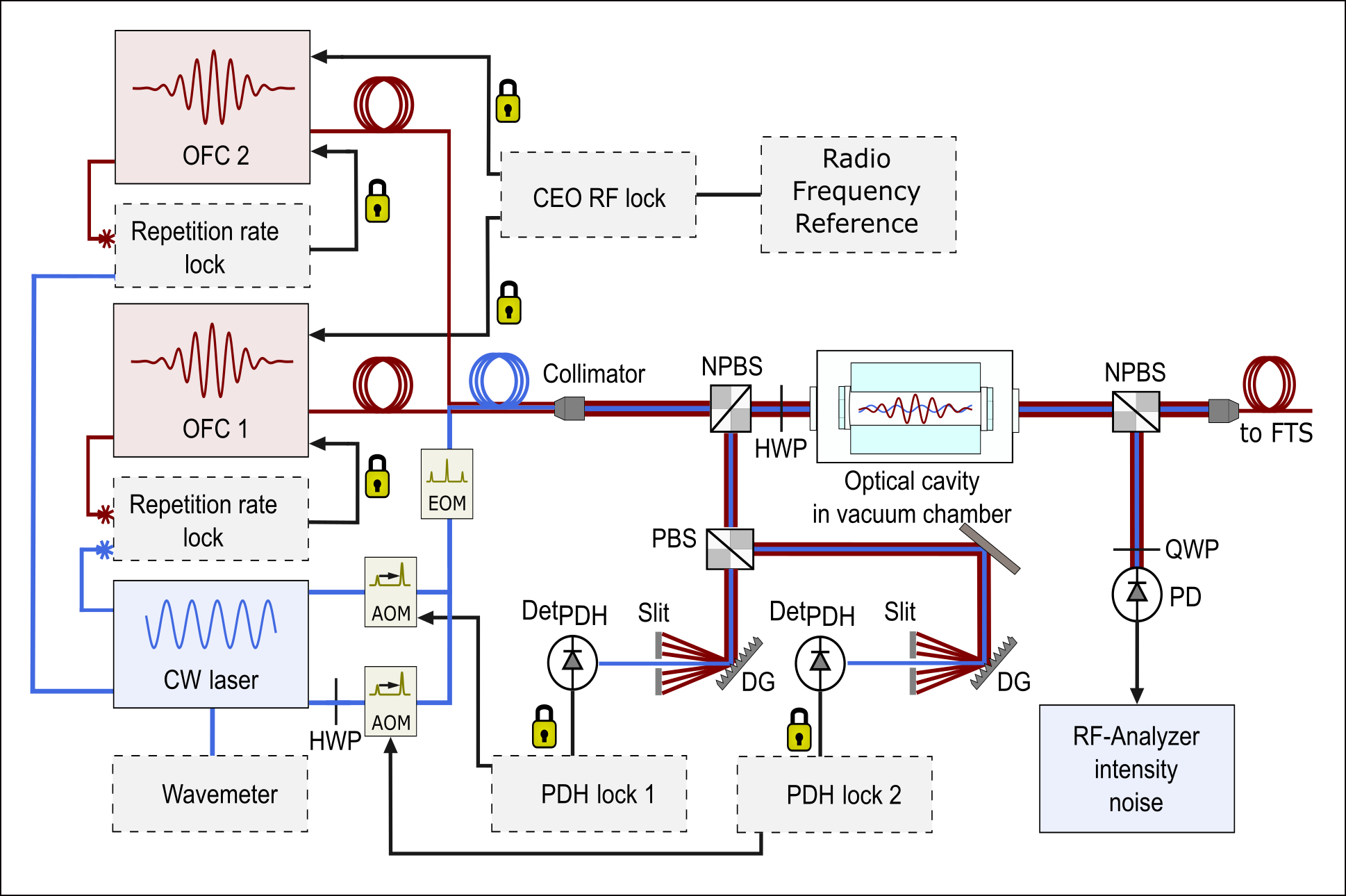}
\caption{Proposed setup scheme. AOM - acousto optic modulator, EOM - electro optic modulator, BS - beam splitter, DG - diffraction grating, Det\textsubscript{PDH} - Detector for PDH-lock, HWP - half-waveplate, QWP - quarter-waveplate, FTS - Fourier transform spectrometer, PD - photo detector. By combing two OFCs with different polarization (e.g. by polarization multiplexing in a polarization-maintaining fiber) and locking both to the same reference laser, the cavity transmission can be controlled for both independently, while at the same time they see the same cavity noise, except the birefringent noise. The cavity transmission can be split, with the beat note between the OFCs incident on a photo diode. Birefringent noise results in intensity noise for different radio frequency beat notes at an RF analyzer.}
\label{fig:Fig6_Proposal}
\end{figure*} 
Previous measurements of birefringent splitting were mostly carried out with CW lasers, usually in the near-infrared wavelength region \cite{cole_tenfold_2013, cole_substrate-transferred_2023, tanioka_study_2023, ma_ultrastable_2024, yang_study_2025, yu_excess_2023, kedar_frequency_2023} with the exception of \cite{winkler_mid-infrared_2021} who focused their work on the mid-infrared. Most of the above extended their research to birefringent noise by measuring the frequency noise of the transmitted probe laser, comparing it to a reference. Among the contributions to this frequency noise are Brownian (mechanical) noise \cite{cole_tenfold_2013}, photo-thermal noise \cite{chalermsongsak_coherent_2016}, thermofringent noise \cite{kryhin_thermorefringent_2023}, photo-birefringent noise as well as intrinsic birefringent noise, and a global polarization-independent noise, which neither depends on mode area nor intracavity power \cite{kedar_frequency_2023, yu_excess_2023}.
While the cancellation of the photo-birefringent effect has been reported \cite{ma_ultrastable_2024, kraus_ultra-stable_2025}, and new frequency stability records have been achieved \cite{lee_frequency_2025}, the origin and behavior of intrinsic birefringent noise remains elusive. 
In order to combine the broadband spectral coverage of the method presented here with intrinsic birefringent noise measurements, we propose a future experiment to record cavity mode dependent birefringent frequency noise.
\\
\\
 The proposed scheme is based on an architecture previously used for the demonstration of dual-comb cavity enhanced spectroscopy presented in \cite{lisak_dual-comb_2022,charczun_dual-comb_2022}.
 Here a CW laser is split in two and one part frequency shifted (e.g. with an AOM) as given in Fig. \ref{fig:Fig6_Proposal}. Their polarization is set to be orthogonal and both are polarization-multiplexed in a polarization-maintaining fiber. The two CW laser outputs (CW~1 and CW~2) are sent to a fiber EOM which creates sidebands for both CW~1 and CW~2. Each is independently locked to a orthogonal eigenmode of an optical cavity. Using a non-polarizing beam splitter (NBPS) together with a polarizing beam splitter (PBS) the reflected light of CW~1 and CW~2 can be split for two independent PDH-locks, with two AOMs acting as fast actuators. With tight locks, and a slow feedback which affects both CW outputs identically, the difference in noise should be negligible.
 
Two optical frequency combs ( OFC~1 and OFC~2) are also polarization multiplexed to the same polarization maintaining fiber as CW~1 and CW~2.  Both OFCs are locked to the corresponding CW laser output, identical to the locking scheme in section \ref{sec:setup_methods}. Since OFC~1 and OFC~2 are locked to the same CW laser they share the same noise, with the exception of the independent PDH-lock. Locking the offset frequencies of both combs to the same value guarantees mutual frequency stability between the two combs.

Directing the transmitted OFCs to a sufficiently fast photo detector, the heterodyne beating between OFC~1 and OFC~2 will result in a down converted RF comb. Due to the Vernier filter, the difference in repetition rate $\Delta f_\mathrm{rep}$ is enhanced to a value $k \times \Delta f_\mathrm{rep}$ as given by Eq. \ref{eq:Vernier}, which becomes the frequency spacing for the down-converted RF comb.
By measuring the beat notes of the down-converted combs we gain access to the difference in frequency noise (FN), as well as intensity noise (IN) of the optical frequencies.

To gain knowledge about the birefringent noise of the cavity modes, investigating the conversion of cavity FN to the individual comb modes is crucial. Following the elastic tape model for self-referenced stabilization \cite{fortier_20_2019}, when passively locking an OFC to a cavity as described in section \ref{sec:setup_methods}, the FN $S_{\mathrm{\nu},n}$ of individual comb modes $n$ is scaled relative to the fixed frequency given by the lock of the comb mode $n_\mathrm{ref}$ to the CW laser
\begin{align}\label{eq:FN_comb_modes}
    S_{\mathrm{\nu},n} = \left( S_\mathrm{\nu, CW} + S_\mathrm{\nu, f_{beat}}\right) \left( \frac{n^2}{n_\mathrm{ref}^2}\right) + S_{\mathrm{\nu}, f_0},
\end{align}
which depends on the FN of the offset frequency $S_{\mathrm{\nu},f_0}$, the FN of the lock given by $S_\mathrm{\nu, f_{beat}}$, as well as the FN given by the CW laser $S_\mathrm{\nu, CW}$, where the latter includes all cavity noise. Investigating the FN of individual beat notes from the down-converted RF comb (i.e. the difference in FN for orthogonal polarizations) only yields information of birefringent noise at the CW laser scaled by a factor $\left( \frac{n^2}{n_\mathrm{ref}^2}\right)$. This holds if the difference in locking noise $\Delta S_\mathrm{\nu, f_{beat}}$ and the difference in offset frequency FN $\Delta S_{\mathrm{\nu},f_0}$ are negligible. Crucially, all three of these quantities can be independently monitored during the measurement .
\\
\\
As mentioned above, we not only have access to the beat note FN, but the IN as well. Any FN of an individual cavity mode $m$ will convert to IN on the individual comb mode incident on said cavity mode. The transmitted intensity $\hat{I}_n$ for one comb mode contains information about the local cavity frequency noise, as well as the birefringent noise at the CW laser, which for small excursion can be expressed in linear form as
\begin{widetext}
\begin{equation}\label{eq:Intensity_comb}
    \hat{I}_n \pm \Delta \hat{I}_n
    = \left( I_{n} \pm \Delta I_{n} \right)
    \left( T\left(\nu_n\right) 
    + \frac{\partial T\left(\nu\right)}{\partial \nu}\big|_{\nu_n} \Delta \nu^\mathrm{comb}_n
    + \frac{\partial T\left(\nu\right)}{\partial \nu}\big|_{\nu_n} \Delta \nu^\mathrm{cav}_m
    \right),
\end{equation}
\end{widetext}
where $I_{n} \pm \Delta I_{n}$ is the intensity of the transmitted comb mode number $n$ and its fluctuation, and $T\left(\nu\right)$ the frequency transmission function of the cavity. 
The frequency dependent change in cavity transmission $\frac{\partial T\left( \nu\right)}{\partial \nu}$ serves as a link between the cavity frequency fluctuation $\Delta \nu^\mathrm{cav}_m$ and the intensity fluctuations after the cavity $\Delta \hat{I}_n$. As the frequency fluctuation at individual comb modes $\Delta \nu^\mathrm{comb}_n$ for a given averaging time can be determined by the FN in eq.\ref{eq:FN_comb_modes}, we can directly infer the cavity mode FN for a given optical frequency $\nu_n$ and known transmission profile $T\left(\nu\right)$. Detecting the IN of the beat note between two transmitted comb modes, each incident on orthogonal polarization eigenmodes of the cavity, gives access to the difference in cavity FN. If both combs have negligible intensity fluctuation $\Delta I_n$ compared to the fluctuation induced by cavity $\frac{\partial T\left(\nu\right)}{\partial \nu}\big|_{\nu_n} \Delta \nu^\mathrm{cav}_m$, the beat note IN will depend solely on the birefringent frequency noise between the corresponding s- and p-polarized cavity mode.
\\
\\
To summarize, since both OFCs are locked to the same CW laser, they share the same cavity noise, except the birefringent noise. Any additional FN between them can only stem from the separate locks, and any shared transient response that arises due to comb-cavity mismatch gets canceled at the detector.
While shared cavity noise is canceled, remaining birefringent FN at an arbitrary cavity mode (same mode number for s- and p-polarization) is converted to amplitude modulation on the incident comb mode, depending on the coincidence with the Lorentzian transmission profile of the cavity. Birefringent FN would thus couple to IN of individual RF beat notes, which could be measured with state-of-the-art equipment.
Tuning the comb over the cavity modes as described in section \ref{sec:setup_methods} gives access to different parts of the optical spectrum in the RF domain. This ensures coverage of the complete optical spectrum, resulting in fully mapped FN over the mirror transmission range. With this, the optical frequency with lowest intrinsic birefringent noise can be found, mitigating the impact of birefringence in cavity based metrology experiments.
In the future, measurements based on CMDS could be used to further probe frequency dependent static and dynamic birefringence in terms of temperature, intracavity power, or induced changes by electric and magnetic fields.

\begin{figure*}[htbp]
\centering
\includegraphics[width=1\textwidth, height = 0.8\paperheight]{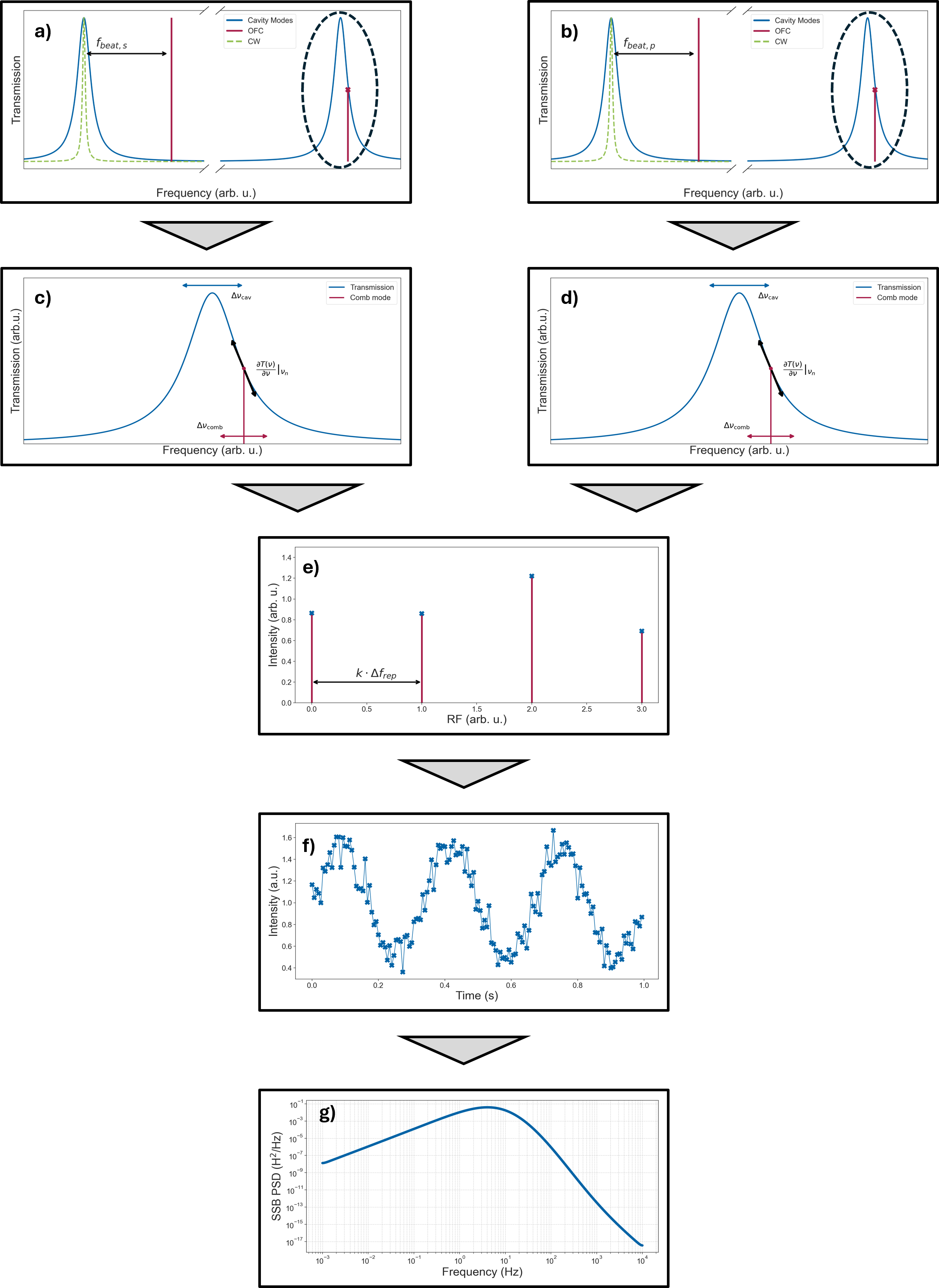}
\caption{Proposed experimental concept and measurement scheme. a), b) CW–OFC locking for s- and p-polarized cavity eigenmodes. An arbitrary number of comb modes coincides with the cavity (e.g. dashed ellipse) such that it is positioned at the steepest part of the cavity transmission profile.   c), d) Conversion of cavity frequency noise (FN) to comb intensity noise (IN) for the two independent polarizations. e) Representation of single-shot down-converted RF spectrum from heterodyne beating of the two OFCs. f) Sketch of a temporal intensity trace of a selected RF beat note. g) Retrieved single-sideband FN power spectral density from f). The scheme enables birefringent cavity FN to be mapped onto measurable RF IN, providing broadband access to birefringent noise across the optical spectrum.}
\label{fig:Fig7_Workflow}
\end{figure*} 

\section{Conclusion}\label{sec:Conclusion}
While high-finesse optical cavities have been used in the past to study minute phase changes in optical coatings, to our knowledge the results presented here depict the first measurement of broadband birefringent mode splitting utilizing an OFC. Additionally, we report the first broadband measurement of birefringent splitting for a crystalline GaAs / AlGaAs cavity. We were able to map the resonance position of individual cavity modes in the stopband with a sensitivity in the order of \SI{5e-14}{} by combining a relative measurement of mode shifts from CMDS with the absolute frequency axis of an OFC. By repeating the measurement for s- and p-polarization, relative to the laboratory reference frame, we were able to calculate the static birefringent splitting on the level of $\Delta n_\mathrm{bir}\approx \SI{300}{ppm}$ over approximately \SI{30}{nm} with a measurement sensitivity compared to optical frequencies on the order of \SI{1e-11}{}. 
Furthermore, by fitting Lorentzian line-shapes to the individual cavity modes, we were able to retrieve individual cavity mode positions and widths. From this we infer that there is no significant difference between the transmission and losses of orthogonal polarizations. Our results show the suitability of CMDS for further investigation of birefringence, not only in the case of crystalline mirrors, but also novel interference coatings and general thin film designs that can be exploited in a cavity structure.
Additionally, we propose a versatile measurement of birefringent noise, which in principle could be used to study different parameters that influence the individual cavity mode FN, e.g. temperature, intracavity power, the influence of incident electric and magnetic fields. Simultaneously probing both polarization eigenmodes could lead to high fractional frequency stability, while accessing Fourier frequencies up to the kHz domain based on the difference in repetition rate for the probe combs. Synchronized CMDS measurements would allow for precise knowledge of the cavity transmission, over broadband acquisition. In conclusion, characterizing and potentially canceling birefringent and the above mentioned noise sources in current and future high-reflective thin film coatings will further enhance modern metrology and push it to ever lower fractional frequency stability.
\begin{acknowledgments}
We thank Lukas W. Perner, Vito F. Pecile, Tom Jungnickel and Garrett D. Cole for valuable discussion.
 For open access purposes, the author has applied a CC BY public copyright license to any author accepted manuscript version arising from this submission.  
This research was funded in whole or in part by the Austrian Science Fund (FWF) [DOI: 10.55776/P36040]. The financial support by the Austrian Federal Ministry for Digital and Economic Affairs, the National Foundation for Research, Technology and Development and the Christian Doppler Research Association is gratefully acknowledged.
\end{acknowledgments}

\section*{Data Availability Statement}
The data that support the findings of
this study are available from the
corresponding author upon reasonable
request.

\bibliographystyle{aipnum4-1}
\nocite{*}
\bibliography{BirefringentBibliography}

\end{document}